\documentstyle[aps,preprint]{revtex} 
\begin{document}

\def\half{{1 \over 2}}

\def\nfdf{${\cal N}=4, d=4$}
\def\ymt{Yang-Mills theory}
\def\sun{SU($N$)}
\def\laa{\langle\kern-.3em \langle}
\def\raa{\rangle\kern-.3em \rangle}
\def\la{\lambda}
\def\cn{{\cal N}}\def\cz{{\cal Z}}
\def\ymt{Yang-Mills theory}
\def\sftH{string field theory Hamiltonian}
\def\aaa#1{a^{\dagger}_{#1}}
\def\aap#1#2{(a^{\dagger}_{#1})^{#2}}
\def\ee{\hbox{e}}
\def\dd{\hbox{d}}\def\DD{\hbox{D}}
\def\tr{\hbox{tr}}
\def\Tr{\hbox{Tr}}
\def\part{\partial}
\def\dj#1{{\delta\over{\delta J^{#1}}}}
\def\cO{{\cal O}}
\def\ssc{\scriptscriptstyle}

\title{Nonperturbative effects in deformation quantization}
 
\author{Vipul Periwal}

\address{Department of Physics, 
Joseph Henry Laboratories,
Princeton University, 
Princeton, NJ 08544 
\tt vipul@princeton.edu}

\maketitle 
\begin{abstract}
The Cattaneo-Felder path integral form of the perturbative Kontsevich
deformation quantization formula is used to explicitly demonstrate
the existence of nonperturbative corrections to na\"\i ve deformation
quantization.
\end{abstract}
\bigskip
\def\st#1#2{#1\star#2}
\def\poi#1#2{\{#1,#2\}}
\def\cf{Cattaneo-Felder }
\def\sw{Seiberg-Witten }
\tightenlines
The physical 
context of the formal problem of 
deformation quantization is the original one set out by 
Dirac\cite{d}\ in making the substitution 
\begin{equation}
\poi fg \rightarrow {1\over i\hbar}
\left(\st{\hat f}{\hat g} - \st{\hat  g}{\hat f}\right)
\end{equation}
the basis for relating classical mechanics on a phase space $M$ to
quantum mechanics, with $\hat f, \hat g$ operators acting on a Hilbert space
of wavefunctions.

In mathematical terms given a Poisson structure on a manifold $M,$ 
the problem\cite{stern}\
is to find an associative product $\star$ on 
the space of formal power series in $\hbar$ with coefficients in
the space of smooth functions on $M$ such that 
\begin{equation}
\st{f}{g} = fg + {i\hbar\over 2}\poi {f}{g} + O(\hbar^{2})
\label{star}
\end{equation}
where $\poi{f}{g}$ is the Poisson bracket on $M.$  
Kontsevich\cite{k}\ gave a solution to this deformation problem
in terms of a formal power series organized as a sum over graphs.
The details of his construction will not be important for us.  

What is more relevant for physics is the 
reformulation of his construction found by Cattaneo and 
Felder\cite{cf}, who gave a path integral form of the Kontsevich 
formula. 
Recall now the Cattaneo-Felder construction:  Let $M$ be a Poisson 
manifold with a Poisson bracket given locally by 
\begin{equation}
\poi{f}{g} = \sum_{i,j=1}^{d} \alpha^{ij}\part_{i}f\part_{j}g
\end{equation}
with $\alpha$ a section of $\wedge^{2}T_{*}M$ satisfying the 
Jacobi identity
\begin{equation}
\alpha^{il}\part_{l}\alpha^{jk} + \hbox{cyclic} \ = 0.
\end{equation}
Let $X$ be a map from the unit disk $D^{2} \rightarrow M$ and 
let $\eta$ be a section of $X^{*}T^{*}M\otimes T^{*}D^{2}.$  Then the
action 
\begin{equation}
S\equiv \int_{D^{2}} \eta_{i}\wedge \dd X^{i} + {1\over 2} 
\alpha^{ij}\eta_{i}\wedge\eta_{j}
\end{equation}
can be used to define a functional integral
such that 
\begin{equation}
\st{f}{g}(x) = \int_{X(1)=x} \DD X\DD \eta \exp(iS/\hbar) f(X(-1))g(X(i))
\label{pois}
\end{equation}
where the integral is over all maps $X:D^{2}\rightarrow M$ such that 
$X(1) = x.$   

When the Poisson structure is associated with a symplectic form $\omega$
the \cf formula simplifies considerably.  The field $\eta$ can be 
integrated out and one is left with
\begin{equation}
\st{f}{g}(x) = \int_{X(1)=x} \DD X 
\exp\left(i\int_{X(\part D^{2})} \dd^{-1}\omega/\hbar\right) f(X(-1))g(X(i))
\label{sympstar}
\end{equation}
where the path integral is now over maps from the circle $S^{1}$ regarded as 
the boundary of the disk $D^{2}$ to the symplectic manifold $M.$  
Since $\omega$ is closed by
definition, it can be represented locally as a one-form denoted
symbolically as $\dd^{-1}\omega.$  In local Darboux coordinates the
integral over the boundary is $S = \int p\dd q,$ therefore the
equations of motion imply that the boundary value is locally 
constant.  In other words, the classical equation of motion maps the
boundary of $D^{2}$ to a point in $M.$  This fact will be important 
for us in the following.  

As in any aspect of quantum physics, the path integral  
is more fundamental than its perturbative saddle-point evaluation, so it 
is appropriate to investigate the Cattaneo-Felder path integral in 
detail to understand its physical content.  
I aim to demonstrate here that there are nonperturbative contributions to 
the Cattaneo-Felder path integral.  These come from topologically 
nontrivial configurations, and hence have coefficients of the
form $\exp(ic/\hbar).$  Since these contributions appear as essential 
singularities in a formal expansion in powers of $\hbar,$  the 
nonperturbative deformation is still a solution to the formal 
deformation problem in eq.~\ref{star}.  

When will there be nontrivial solutions to the classical equations of 
motion?  We wish to evaluate a sum over maps from the disk $D^{2}$ 
to the symplectic manifold $M$ such that the boundary of the disk $S^{1}$
is mapped to the point $x$ in $M.$  In general, homotopy classes of 
maps from the  $n$-disk $D^{n}$ 
to a manifold $M$ relative to a submanifold $N$ of $M$ which map the
boundary of $D^{n}$ to $N$ are elements in the relative homotopy group
$\pi_{n}(M,N).$  (Relative and absolute 
homotopy groups are defined with a choice 
of basepoint, but we have not made this dependence explicit
in our notation.)  In our case, we evidently need to consider 
$\pi_{2}(M,N\equiv\{x\}),$ but this   is isomorphic
to the absolute homotopy group $\pi_{2}(M).$

We expect then that we will get nonperturbative contributions to the 
path integral in cases where $M$ has non-vanishing $\pi_{2}.$  For 
example, $\pi_{2}(S^{2}) = {\bf Z}$ and $\pi_{2}(T^{2n}) = 0$ so there 
should be such contributions for $S^{2}$ and no such contributions for 
$T^{2n}.$

In the simplest case $M=S^{2},$ the homotopy classes of maps are
just classified by the degree of the map.  
It is then easy to evaluate the action for degree $n$ 
solutions to the classical 
equations of motion $X_{n}$ since
\begin{equation}
\hbox{deg}(X_{n}) \int_{S^{2}} \omega = \int_{D^{2}} X_{n}^{*}\omega \equiv S
\end{equation}
and $V\equiv \int_{S^{2}} \omega$ is just the symplectic volume of $S^{2}.$
Notice that the value of the action of $X_{n}$ does not depend on the 
detailed form of $X_{n},$ just on the topological class given by the 
degree $n.$  

Thus the \cf  path integral evaluated semiclassically for $M=S^{2}$ is
\begin{equation}
\sum_{n\in {\bf Z}} \exp(inV/\hbar)  \langle 
f(X_{n}+\xi(-1))g(X_{n}+\xi(i))\rangle_{n}
\label{sumoverinstantons}
\end{equation}
where  $\langle \ldots\rangle_{n}$ 
denotes the expectation value in the path integral 
evaluated perturbatively about $X_{n},$  with $\xi$ the fluctuation.
Thus we see 
contributions with essential singularities as functions of the 
deformation parameter $\hbar$ from topologically nontrivial sectors of the 
path integral.  

I should add a few words on the perturbative evaluation of the path 
integral about these solutions.  The analysis here is 
restricted to symplectic manifolds and one does not need all the
sophistication necessary for the general case\cite{cf}.  
The gauge symmetry of the model in this case 
is diffeomorphism invariance since the action does not depend 
on any metric on the disk.   If we consider K\"ahler 
manifolds, then a natural gauge fixing would be to localize on 
holomorphic representatives in each topological sector.  This can be 
done in a straightforward fashion with a small modification of Witten's work on 
topological sigma models\cite{wtopsigma}.  The standard formal path 
integral argument for associativity of the product is unchanged
of course.

I want to emphasize that there is no shortcoming from a mathematical 
point of view in the work of Kontsevich\cite{k}.  His formula is 
perfectly adequate as a {\it formal} deformation quantization, but the 
importance of nonperturbative contributions cannot be over-emphasized 
from the physical perspective.  Indeed,
sums over such topologically nontrivial configurations  are crucial 
for the appearance of mirror symmetry for example.  There is no 
obvious duality symmetry that one expects in the simple example of 
$M=S^{2}$
considered here, but in general any $T$-duality like symmetry would 
require inclusion of such topologically non-trivial configurations in 
the path integral.  $\pi_{2}$ is nontrivial for any simply connected
Calabi-Yau manifold, for example. 

I do not see immediately 
how to carry out this argument in the general case of a
Poisson manifold (eq.~\ref{pois}) 
since the local constancy of the boundary classical 
configuration was crucial in identifying nontrivial configurations 
with elements in the homotopy group.  The  
Seiberg-Witten\cite{sw}\ limit presumably requires the general case.

Acknowledgements: I thank M. Berkooz, D. Gross,
G. Lifschytz, D. Morrison, R. Schimmrigk,
P. Schupp and E. Verlinde for helpful discussions.
This work was supported in part by NSF grant PHY98-02484.

\def\np#1#2#3{Nucl. Phys. B#1,  #3 (#2)}
\def\prd#1#2#3{Phys. Rev. D#1, #3 (#2)}
\def\prl#1#2#3{Phys. Rev. Lett. #1, #3 (#2)}
\def\pl#1#2#3{Phys. Lett. B#1, #3 (#2)}


\begin{thebibliography}{99}

\bibitem{d} P.A.M. Dirac, {\it The Principles of Quantum Mechanics} 
(Oxford U. Press, Fourth ed., 1958)

\bibitem{stern} For a recent review, see D. Sternheimer, {\sl 
Deformation quantization: Twenty years after}, math/9809056

\bibitem{k} M. Kontsevich, {\sl Deformation quantization of Poisson 
manifolds}, q-alg/9709040

\bibitem{cf} A.S. Cattaneo and G. Felder, {\sl A path integral 
approach to the Kontsevich quantization formula}, math.QA/9902090

\bibitem{wtopsigma} E. Witten, Comm. Math. Phys. {\bf 118}, 411 (1988)

\bibitem{sw} N. Seiberg and E. Witten, J. High Energy Phys. {\bf 
9909}, 032 (1999)

\end{thebibliography}
\end{document}